\documentclass[aps,prl,reprint,groupedaddress]{revtex4}
\usepackage{mathrsfs}
\usepackage{cancel}
\usepackage{amssymb}
\usepackage{comment}
\usepackage{graphicx}
\def\a{\alpha}
\def\b{\beta}

\def\p{\partial}  
\def\ep{\epsilon}  
\def\o{\over}
\def\p{\partial}
\def\l{\left}
\def\r{\right}

\def\U5{\tilde U_5}
\def\be{\begin{equation}}
\def\ee{\end{equation}}
\def\bea{\begin{eqnarray}}
\def\eea{\end{eqnarray}}
\def\f2{F^{\a\b}F_{\a\b}}
\def\16gc2{{16\pi G\over 3c^2}}

\begin{document}


\title{Maxwellian mirages in general relativity}




\author{L.L.~Williams}
\email{willi@konfluence.org}
\homepage[www.konfluence.org]{}
\affiliation{Konfluence Research Institute,\\ Manitou Springs, Colorado}

\author{N. Inan}
\email{ninan@ucmerced.edu}
\affiliation{Clovis Community College,\\ Clovis, California}


\date{30 March 2021}
\begin{abstract}
Maxwellian approximations to linear general relativity are revisited in light of relatively recent results on the degrees of freedom in the linear gravitational field. The well-known Maxwellian formalism obtained in harmonic coordinates is compared with a Maxwellian formalism obtained under a coordinate choice where each of the metric components corresponds to each of the coordinate-invariant degrees of freedom of the linear gravitational field. The coordinate freedom of general relativity can be exploited to cast the field equations into Maxwellian form, but such forms can be mere mirages of the coordinate choice -- mirages such as vector gravitational waves. A coordinate choice that yields perfectly-Maxwellian field equations, will yield a force equation that is not Lorentzian. If field definitions are chosen to obtain Lorentz-like terms in the force equation, then Maxwellian forms are compromised in the field equations. Many treatments of gravito-electromagnetism will make inconsistent ordering choices between the field equations and force equations, or else truncate terms of relevant order from the force equation. Often such mistakes reflect an attempt to force exact Maxwellian analogs simultaneously in both the field equations and the force equation, with the result that terms dropped are as large as those kept.

\end{abstract}


\maketitle


\section{1. Introduction}

Considerations of relativity imply there are only 3 basic mathematical types of long-range force fields in nature: scalar fields, vector fields, and tensor fields. These 3 fundamental field types are depicted in Figure 1. We hope nature is not so malicious as to prescribe a third-rank tensor field in nature. 

A scalar field has a single potential; a vector field has 4 potentials; and a symmetric tensor field has 10 potentials. So far, 2 of the 3 long-range fields are identified in nature. The classical electromagnetic field is a vector field, because its potentials are the 4 components of a 4-vector, $A^\mu$. The gravitational field is a tensor field, because its potentials are the 10 components of a 4D tensor, $g_{\mu\nu}$. The complexity of the gravitational field harbors electromagnetic-like behavior under certain approximations and certain coordinate choices, inviting a vector approximation to the tensor potentials, with vector field equations instead of tensor field equations. There is no long-range scalar field identified in nature. 

\begin{figure*}
\includegraphics[width=1.0\textwidth]{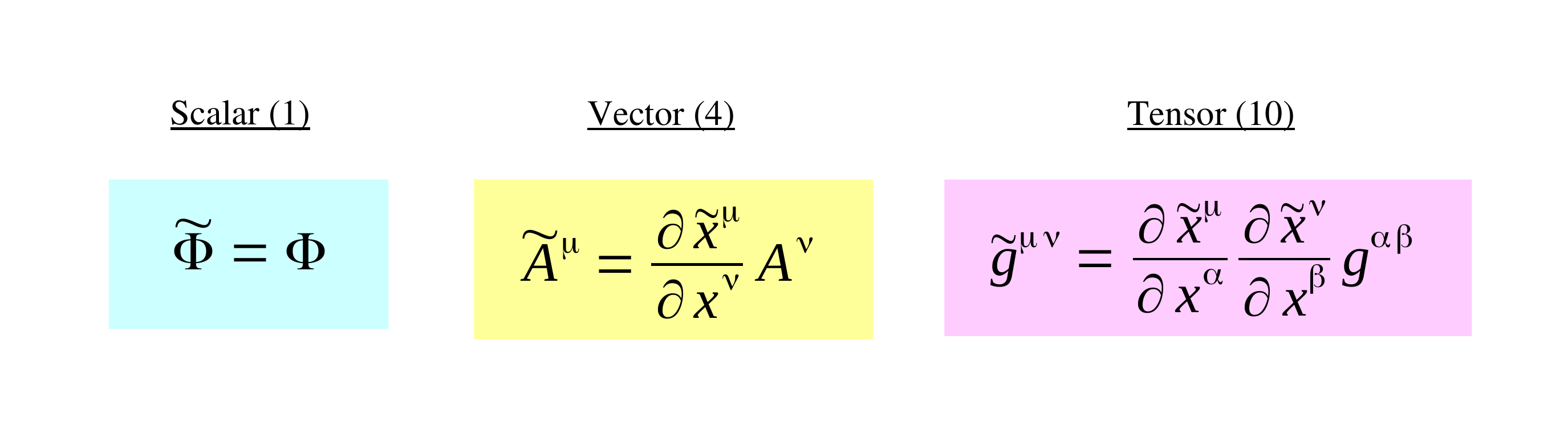}
\caption{Three basic types of classical field, according to how they behave under a coordinate transformation between coordinates $x^\mu$ and ${\widetilde x}^\mu$. There are always 4 coordinates, corresponding to the 4 dimensions of space and time, and summation over coordinates is implied on repeated indexes. The classical fields are long-range and their bosons are massless. A scalar field has no indexes, 1 component, and is invariant. A vector field has 1 index, 4 components, and transforms linearly in the transformation matrix. A symmetric tensor field has 2 indexes, 10 components, and transforms quadratically in the transformation matrix. The classical electromagnetic field is a vector field, and gravity is a tensor field. No long-range scalar field has been found.\label{Fig2}}
\end{figure*}

This paper addresses use of vector gravity approximations of, or substitutions for, the field equations of general relativity. By ``vector gravity", we refer to a theory of gravity that describes the gravitational field with the 4 potentials of a vector field, instead of the 10 potentials of a tensor field. The result is a set of field equations that look like the Maxwell equations, and/or a force equation that looks like the Lorentz force law, so we also call this ``Maxwellian gravity." 

There are various semantics for what we are calling ``vector gravity" and ``Maxwelllian gravity". Others may use terms like ``gravito-electromagnetism". Depending on how a given author defines the term ``gravito-electromagnetism", it may be well-founded or not. There is no clear and widely accepted mathematical definition of the term, and to some extent, it is in the eye of the beholder. The term has also been applied to more-sophisticated considerations than the ones given here. We shall use the term for obvious identifications in the force equations or the field equations.

Maxwellian vector gravity was first introduced by Heaviside \cite{heav} in 1893. Yet it came into widespread use after the discovery of the Einstein field equations in 1915, as researchers sought approximations to the complexity of tensor gravity. Two prominent works on vector gravity came in the mid-20th century, Sciama \cite{sciama} in 1953, and Forward \cite{forward} in 1961. Although influential, those two papers did not proceed from systematic approximations of general relativity, arguing instead mainly by analogy from electromagnetism.

The vector-like nature of linear gravity was put on a firmer mathematical footing by Bragisnky et al.\cite{bct} in 1977. Those authors arrived at a Maxwellian description of gravity based on a Parameterized Post Newtonian analysis, in turn based on Newtonian-limit results in the famous textbook by Misner, Thorne, and Wheeler.\cite{mtw} 

Harris \cite{harris} provided an analysis much later of the Maxwellian equations emerging in harmonic coordinates. Harris got different results in the Maxwellian equations than Braginsky et al. regarding the gravitational Faraday law.

Since those classic results were established, the degrees of freedom in the linear gravitational field have become better understood. Flanagan \& Hughes \cite{fh} obtained the coordinate-invariant degrees of freedom of the linear gravitational field. Their treatment was adopted by Poisson \& Will \cite{pw}, and we shall apply those results to consider the relation between Maxwellian gravitational fields and gravitational degrees of freedom.

It seems relatively common for gravitational researchers to conflate electromagnetic behavior in the gravitational field with the Maxwellian forms emerging in linear general relativity. A common error is to infer 3-vector gravito-electromagnetic waves similar to those found in the Maxwell equations. Another is to to order the field equations and force equation inconsistently, in order to obtain exact Maxwellian analogs in both the field equations and in the force equation.

The utility of linear approximations to general relativity can be limited. Yet they are often utilized without a careful consideration of their physical suitability. There are several fundamental distinctions between gravitational and electromagnetic fields that should be considered prior to any linear approximation to GR.

The main conceptual problem of an electromagnetic-like approximation to gravity is to approximate 10 tensor potentials with some combination of 4 potentials, or with 4 combinations of the 10 potentials. The tensor nature of the gravitational field enters at zeroth order in source velocity, as can be seen in the Newtonian line element in isotropic form:
\be
\label{nle}
ds^2 = -(1+2\phi)c^2 dt^2 + (1-2\phi)(dx^2 + dy^2 + dz^2)
\ee
The gravitational field cannot be represented by a vector field because a vector field does not have the spatial tensor components for the line element (\ref{nle}). Those components play a role in perihelion precession and bending of light.

The non-linearity itself is a key difference between gravity and electromagnetism. The non-linearity arises because gravitational fields themselves create gravitational fields, since gravity couples to all energy forms. By comparison, the electromagnetic field is not electrically-charged, and so electric forces superpose. 

This means vector gravity must be only a weak-field limit of the tensor theory, where non-linear effects can be ignored. Under the linear assumption, the gravitational field transfers no energy to gravitational sources, thereby eliminating self-consistent matter-and-gravity coupling (back-reaction) from the domain of vector approximations.

A vector approximation to gravity vastly undercounts the complexity of the gravitational forces. The gravitational force $\Gamma^\mu_{\a\b}$ has 40 components. Whereas, the vector gravito-electric and -magnetic forces have only 6 components. Vector gravity leaves a great deal of gravitational force complexity unaccounted for.

Nonetheless, vector gravity does offer some insight into the complexities of tensor gravity. In certain parameter regimes, such as in frame dragging, the gravitomagnetic effects can predominate. Yet coordinate freedom can lead unwary researchers astray, as we will see below.

Our approach is to look for electromagnetic-type force fields in the force equations, and then follow them into the field equations. We find that the apparent gravito-electromagnetic fields that emerge naturally in the gravitational force equation do not emerge naturally in the field equations. We discuss the coordinate-independent degrees of freedom of the linear gravitational field, and how coordinate information can become encoded in the potentials in general coordinate systems. While gravity has electromagnetic-type force effects from the 4 time components of the metric perturbation, the gravito-electromagnetic 3-forces are coordinate-dependent. 

We will then revisit how harmonic coordinates can be used to obtain Maxwell-like equations in terms of gravito-electric and gravito-magnetic fields. We will compare those results with an alternative set of Maxell-like gravito-electric and -magnetic fields written in transverse coordinates, that better represent the physical degrees of freedom of the linear gravitational field. We then provide a retrospective discussion of the papers by Sciama, Forward, and Harris.

\section{2. Review of Tensor and Vector Fields}
Let us review the tensor field equations of gravitation, and the vector field equations of electromagnetism; and also the force equations of gravitation and electromagnetism. These are standard textbook results to establish the basis of the discussion. We pay particular attention to the constraints on the field equations implied by conservation laws, and the resulting implication for degrees of freedom in the fields.

The gravitational field is characterized by a symmetric tensor potential $g_{\mu\nu}$, the metric tensor, where the greek indices range over the 4 coordinates of space and time. There are thus 10 components of the metric. The gravitational field equations for these components are given by the Einstein equations
\be
\label{EE}
G_{\mu\nu} = {8\pi G \o c^4}T_{\mu\nu}
\ee
where $G_{\mu\nu}$ is the Einstein tensor, containing non-linear derivatives of the metric up to second order; $G$ is the gravitational constant; $c$ is the speed of light; and $T_{\mu\nu}$ is the energy-momentum tensor of matter and radiation. 

The Einstein equations (\ref{EE}) obey a set of constraints informally known as the Bianchi identities (technically, they are the twice-contracted Bianchi identities):
\be
\label{BI}
\nabla_{\mu}G^{\mu\nu} \equiv 0 = \nabla_{\mu}T^{\mu\nu}
\ee
where $\nabla_{\mu}$ is the covariant derivative for the metric.

Equation (\ref{BI}) expresses both an identity of the Einstein tensor, and conservation of energy-momentum of the matter and radiation sources. The Einstein equations (\ref{EE}) give 10 equations in 10 unknowns of the metric, yet the Bianchi identities (\ref{BI}) provide 4 constraints on the Einstein equations, so that there are only 6 independent equations in the Einstein equations. The extra 4 equations necessary to close the system are the coordinate choice, and this under-determinedness is an expression of general covariance. The coordinate choice itself is not covariant.\cite{weinberg} The approximations to the Einstein equations, because they select a coordinate system, are not themselves covariant. The covariance manifests only in the Einstein equations, not their applications and approximations.

When the energy-momentum comprises a cold gas of non-interacting particles with no internal degrees of freedom, conservation of energy-momentum (\ref{BI}) reduces to the geodesic equation:
\be
\label{GE}
U^\mu \nabla_{\mu} U^\nu = {dU^\mu\o d\tau} + \Gamma^\mu_{\alpha\beta} U^\alpha U^\beta = 0
\ee
where $U^\mu\equiv dx^\mu/d\tau$ is the 4-velocity of a particle; $\Gamma^\nu_{\alpha\beta}$ is the affine connection, which depends on first derivatives of the metric, and constitutes the gravitational force; and $\tau$ is the proper time given by $c^2 d\tau^2 = g_{\mu\nu} dx^\mu dx^\nu$. 

Now consider the case of the electromagnetic field in flat spacetime, a vector field whose potential is a covariant four-vector, $A_\mu$. The field strength tensor is given by $F_{\mu\nu} \equiv \p_\mu A_\nu - \p_\nu A_\mu$, and the field equations, in this case the Maxwell equations, are given by
\be
\label{ME}
\p_\mu F^{\mu\nu} = \mu_0 J^\nu
\ee
where $\mu_0$ is the permeability of free space and $J^\nu$ is the electromagnetic current 4-vector.

The Maxwell equations obey a constraint from conservation of charge that corresponds to an identity of the field strength tensor:
\be
\label{conschg}
\p_\mu \p_\nu F^{\mu\nu} \equiv 0 = \p_\nu J^\nu
\ee

The Maxwell equations (\ref{ME}) are 4 equations in the 4 potentials of $A^\mu$, but the constraint (\ref{conschg}) on $A^\mu$ implies the Maxwell equations provide only 3 independent equations in the 4 unknowns. The 4th equation is provided by an additional, freely-chosen constraint on the potentials called the ``gauge".

The Lorentz force law describes the effect of electromagnetic forces on a body of electric charge $q$ and rest mass $m$:
\be
\label{LFL}
{dU^\mu\o d\tau} = {q\o m} F^{\mu\nu}U_\nu
\ee
This force equation is independent of the choice of electromagnetic gauge. There is a freedom in the choice of the potentials, ``gauge freedom", that leaves the forces in (\ref{LFL}) unchanged. 

Additional insight is gained by considering the traditional 3-vector electric and magnetic force fields. To reinforce Maxwellian identifications, let us adopt bold notation for the components of a 3-vector. Therefore, ${\bf E} \equiv E^i$, where small roman indexes $i,j,k$ span the 3 spatial coordinates. We can similarly write the spatial components of the potential 4-vector as $A^\mu = (\phi /c , A^i) \equiv (\phi /c, \bold A)$. Let us also employ bold notation for 3-space curls and divergences, such that ${\bf \nabla\times E} \equiv \epsilon_{ijk}\p_j E^k$, and ${\bf \nabla\cdot E} \equiv \p_i E^i$. 

With this notation, let us define the electric and magnetic 3-vector components from the potential 4-vector components:
\be
\label{ebfield}
{\bf E} \equiv - \p_i \phi - \p_t \bold A
\quad , \quad
{\bf B} \equiv \bold{\nabla\times A}
\ee
 
Then the Maxwell equations are expressed
\be
\label{maxB}
\bold{\nabla\cdot B} = 0 \quad,\quad \bold{\nabla\times B} = \mu_0 \bold J + \mu_0 \epsilon_0 \p_t {\bf E}
\ee
\be
\label{maxE}
\bold{\nabla\cdot E} = \rho/ \epsilon_0 \quad,\quad \bold{\nabla\times E} = - \p_t {\bf B}
\ee
where the 4-current is decomposed as $J^\mu = (c \rho, \ \bold J)$, and where $\ep_0$ is the permittivity of free space. Note that Faraday's Law is obtained from the definitions (\ref{ebfield}), and so inductive effects are built into these definitions of the fields.

It is clear from (\ref{maxB}) and (\ref{maxE}) that the 3-vector force fields obey wave equations in the absence of sources:
\be
\label{maxwave}
\nabla^2 {\bf E} = \mu_0 \epsilon_0 {\p^2{\bf E} \o \p t^2}
\quad , \quad \nabla^2 {\bf B} = \mu_0 \epsilon_0 {\p^2{\bf B} \o \p t^2} 
\ee
where we are once more using 3-space vector notation to indicate component derivatives such that $\nabla^2 \equiv \p_i \p^i$.

With a choice of electromagnetic gauge $\p_\mu A^\mu = 0$, the equation for the potentials (\ref{ME}) can also be transformed to a wave equation:
\be
\label{potwave}
\nabla^2 A^\mu - \mu_0 \epsilon_0 {\p^2 A^\mu \o \p t^2} = - \mu_0 J^\mu
\ee

The Lorentz force law (\ref{LFL}) is expressed in terms of 3-forces:
\be
\label{LFL3e}
{d\varepsilon\o dt} = q {\bf E} \cdot \bold v
\ee
\be
\label{LFL3}
{d\bold p \o d t} = q{\bf E} + q \bold{v\times B}
\ee
where $mU^\mu \equiv m(U^t, {\bf U})\equiv  (\varepsilon / c, \bold p) = (\varepsilon /c, \ \varepsilon \bold v /c^2)$, in terms of particle mass-energy $\varepsilon$, speed $\bf v$, and momentum $\bold p$.

Note that the gravitational equations (\ref{EE}), (\ref{GE}), and the electromagnetic equations (\ref{ME}), (\ref{LFL}), (\ref{maxB}), (\ref{maxE}), (\ref{LFL3e}), and (\ref{LFL3}) are all fully relativistic and exact.

\section{3. Consideration of the linear geodesic equation}
The route to Maxwellian patterns in the field equations of general relativity is to presume small perturbations of the metric, so that quadratic and higher powers of the perturbation can be ignored, and the theory can be linearized:
\be
\label{linmetric}
g_{\mu\nu} = \eta_{\mu\nu} + h_{\mu\nu} ~,~ \vert h_{\mu\nu}\vert \ll 1
\ee
This decomposition amounts to a restriction to a limited class of coordinate transformations that obey (\ref{linmetric}). The implied restriction to such coordinates means the linear approximation, like all other approximations to general relativity that select a coordinate system, is not generally covariant.

Let us start our investigation of Maxwellian gravity with a consideration of the geodesic equation (\ref{GE}), to linear order in $h_{\mu\nu}$. We could consider a more-complex energy-momentum tensor in (\ref{BI}), but the geodesic equation allows straightforward considerations of test particles. Then we will follow whatever Maxwellian force patterns emerge into the field equations.

Our analysis will work from more recent developments in gravitational field theory that emphasize the importance of coordinate-independent fields, and will reconsider well-known Maxwellian results in general relativity with these coordinate-independent fields in mind. The key to such an analysis is to recognize how components of $h_{\mu\nu}$ behave under spatial rotations: the components $h_{ti}$ transform as a 3-vector, the component $h_{tt}$ transforms as a scalar, and the components $h_{ij}$ transform as a spatial tensor.\cite{carroll, pw} We will introduce notation for the components of $h_{\mu\nu}$ that recognize these identifications. Our notation allows us to group components of $h_{\mu\nu}$ that transform only into themselves under spatial rotations. Additionally, we must decompose the  trace and trace-free parts of $h_{ij}$, since the spatial trace contains zeroth-order metric perturbations.

Let us follow Ref.\cite{carroll}, and decompose $h_{\mu\nu}$ in these terms:
\be
\label{lintime}
h_{tt} \equiv -2\phi \quad , \quad h_{ti} \equiv w^i \equiv \bold w
\ee
\be
\label{hij}
6\Psi \equiv - \delta^{ij} h_{ij} \quad , \quad 2 s_{ij} \equiv  h_{ij} + 2\Psi\delta_{ij}
\ee
The normalization of $h_{tt}$ anticipates its identification with the Newtonian potential.

The linear metric (\ref{linmetric}) is substituted into the geodesic equation (\ref{GE}) and terms are kept to linear order in the field perturbations. Then we can write separately the time and space components of (\ref{GE}). The time component is given as
\be 
\label{linE}
{1\o \varepsilon}{d\varepsilon\o dt} =  
-\p_t\phi - 2 v^i \p_i\phi 
+{v^i v^j\o 2c}(\p_i w_j + \p_j w_i) 
 +  {v^i v^j\o c^2}(\p_t \Psi - \p_t s_{ij})
\ee
where we have used $dt/d\tau = \varepsilon/mc^2$ to transform to ordinary time derivatives.

The spatial components are
\be
\label{linmomcomp}
{1\o\varepsilon}{d p^i\o dt} = 
-{1\o c}\p_t w^i - \p_i \phi + {v^j\o c} (\p_i w_j - \p_j w_i)
- 2 {v^j\o c^2} \p_t s_{ij} + 2 {v^i\o c^2} \p_t\Psi 
+ 2 {v^i v^j\o c^2} \p_j \Psi - {v^2\o c^2} \p_i\Psi
- {v^j v^k\o c^2} (\p_j s_{ki} + \p_k s_{ij} - \p_i s_{jk}) 
\ee

The spatial components (\ref{linmomcomp}) clearly invite a grouping of terms that looks like (\ref{ebfield}). The force equation therefore justifies the identification of gravito-electric and gravito-magnetic 3-vectors ${\bf E}_g$ and ${\bf B}_g$, such that
\be
\label{gravebfield}
{\bf E}_g \equiv E_g^i \equiv - \p_i \phi - \p_t w^i/c 
\quad , \quad
{\bf B}_g \equiv B_g^i \equiv \epsilon_{ijk} \p_j w^k \equiv \nabla\times {\bf w}
\ee
The units of ${\bf E}_g$ and ${\bf B}_g$ are inverse length, because the $h_{\mu\nu}$ potentials are unitless. In other treatments of Maxwellian gravity \cite{forward, harris,bct}, the electric field is given units of acceleration. Refs.~\cite{bct,harris} assign units of acceleration to the magnetic field; ref.~\cite{forward} assigns units of inverse time. Our expression for ${\bf E}_g$ can be converted to units of acceleration by multiplying by $c^2$. We elect to stay in natural units because we will be comparing gravito-electromagnetic fields between coordinate systems.

Newtonian gravity is seen as a Coulomb-like 3-force emerging from a 4-vector-like $h_{t\mu}$, exactly as electric 3-forces emerge from the electromagnetic 4-vector potential. The gravito-magnetic 3-force emerges from the same $h_{t\mu}$. A gravitational Faraday's law follows from (\ref{gravebfield}) by definition, $\nabla\times{\bf E}_g \equiv -\p_t {\bf B}_g$. These features make a strong analogy between electromagnetic and linear gravitational force effects. The time components of the metric perturbation behave very much like a gravito-electromagnetic vector potential. 

Yet there are other terms in (\ref{linmomcomp}) whose effects arise from the tensor nature of the gravitational field, $h_{ij}$, and which have no electromagnetic analogy. They are mathematically complex compared to the electromagnetic-like terms. The most important are the terms in $\Psi$, arising from the spatial trace. Metric perturbations in the spatial trace exist at Newtonian order in the gravitational field. While electromagnetic-like force effects emerge naturally in the linear geodesic equation, the tensor effects may also be important and cannot be ignored. Therefore, let us define another 3-force field ${\bf N}_g$ that represents lowest-order tensor effects in the gravitational field:
\be
{\bf N}_g \equiv N_g^i \equiv -\p_i \Psi 
\ee
To distinguish it from the {\it electric field} ${\bf E}$ and the {\it magnetic field} ${\bf B}$, let us call ${\bf N}_g$ the {\it neutral field}.

With these definitions, let us recast (\ref{linmomcomp})
\be
\label{linmom}
{1\o\varepsilon}{d p^i\o dt} = 
{E}^i_g + \epsilon_{ijk}{v^j\o c} B^k  
+{1\o c^2} N_g^j (v^2 \delta^i_j - 2v^i v_j)
+2 {v^i\o c^2} \p_t\Psi 
- 2 {v^j\o c^2} \p_t s_{ij}
- {v^j v^k\o c^2} (\p_j s_{ki} + \p_k s_{ij} - \p_i s_{jk}) 
\ee
Note these equations are fully relativistic, and that the equations for $\varepsilon$ and $\bf p$ are not separated. Much of the complexity of the tensor effects can be removed by ignoring terms quadratic in particle speed. The dominant part of $\varepsilon$ is the rest mass energy $mc^2$, but it also contains terms quadratic in speed, and linear in the metric perturbations. Let us therefore simplify (\ref{linE}) and (\ref{linmom}) by multiplying through by $\varepsilon$ and writing them to linear order:
\be 
\label{linE1}
{d\varepsilon\o dt} =  
-mc^2 (\p_t\phi + 2 v^i \p_i\phi ) + O(v^2/c^2)
\ee
\be
\label{linmom1}
{d {\bf p}\o dt} =
m({\bf E}_g c^2 + {{\bf v}} \times {\bf B}_g c
+ {2{\bf v}}\p_t\Psi 
- {2v^j}\p_t s_{ij}) + O(v^2/c^2)
\ee

In these expressions, we are now mixing bold and index notation in the same equation, so that the set of 3 components indicated by the bold symbol map to the set of 3 components indicated by a small roman index. We stay with tensor component notation, but allow bold notation for vector components in addition to small roman indices.

It seems clear from (\ref{linmom}) that there are ``gravito-magnetic" forces, to the extent that they functionally depend on ``$\bold{v\times} {\bf B}$". In fact, these are the forces of frame-dragging or the Lens-Thirring effect. We are also free to identify Newton's law with ${\bf E}_g$ in time-independent fields, and the electromagnetic analogy seems complete. Gravito-electric and gravito-magnetic 3-forces emerge from the 4 $h_{t\mu}$ potentials, but other force terms, perhaps equally important, emerge from $h_{ij}$. 

The 4 time components $h_{t\mu}$ approximate a 4-vector in their transformation properties.
A vector is distinguished by its coordinate transformation, $A_\nu \rightarrow A_\nu'$ as in Figure 1:
\be
A_\nu' = {\p {x}^\mu \o \p {x'}^\nu} A_\mu
\ee
Treating the $h_{t\mu}$ as components of a tensor, then
\be
\label{tmet}
{h'}_{t\mu} = {\p {x}^\a \o c\p t'}{\p {x}^\b \o \p {x'}^\mu} h_{\a\b} \simeq 
{\p {t} \o \p t'}{\p {x}^\b \o \p {x'}^\mu} h_{t\b}
\simeq {\p {x}^\b \o \p {x'}^\mu} h_{t\b}
\ee
where the approximation follows for non-relativistic Lorentz transformations $\p x^i/c\p t \ll 1$.
We see in (\ref{tmet}) that $h_{t\mu}$ approximates a 4-vector, and the electromagnetic analogy is reinforced. Be careful to note that the $h_{\mu\nu}$ do not obey the strict definition of a tensor, because they exist in a restricted class of coordinate transformations that preserve the Minkowski metric at zeroth order. Nonetheless, the $h_{t\mu}$ do behave as a tensor under Lorentz transformations.

For time-independent gravitational fields, the geodesic equation (\ref{linmom1}) reduces to an equation that is identical with the Lorentz force law of electromagnetism. But this identification is only true for weak, time-independent gravitational fields and for non-relativistic particles. If there are time-dependent gravitational fields, then the tensor terms in the force equation can be as important as the gravito-magnetic term. Nonetheless, let us follow the thread of (\ref{gravebfield}) from the geodesic equation, into the field equations, and see if and how Maxwellian field equations arise.

\section{4. Consideration of the linear field equations}

The linear gravitational field equations are obtained by using (\ref{linmetric}) in (\ref{EE}) and keeping terms in $G_{\mu\nu}$ to linear order in the perturbations. Even this simple linear approximation to the Einstein equations, however, is quite complicated. Let us follow \cite{carroll} and write the components of the linear Einstein tensor:
\be
\label{Gtt}
G_{tt} = 2 \nabla^2 \Psi + \p_{ij} s^{ij}
\ee
\be
\label{Gtj}
G_{tj} = - {1\o 2} \nabla^2 w_j + {1\o 2} \p_{jk} w^k + {2\o c}\p_{tj} \Psi + {1\o c}\p_{tk} s_j^k
\ee

\be
\label{Gij}
G_{ij} = (\delta_{ij} \nabla^2 - \p_{ij})(\phi - \Psi) + \delta_{ij} \l[ {1\o c}\p_{tk} w^k + {2\o c^2} \p_{tt} \Psi -  \p_{km} s^{km} \r]
 - {1\o 2c}(\p_{ti} w_j + \p_{tj} w_i)  - 
\Box s_{ij} + \p_{ki} s_j^k + \p_{kj} s_i^k 
\ee
where $\Box\equiv \eta^{\mu\nu}\p_\mu \p_\nu$.

There are several important conclusions to draw from the linear Einstein tensors (\ref{Gtt}, \ref{Gtj}, \ref{Gij}), forming the left hand side of (\ref{EE}). One is that the ${\bf E}_g$ and ${\bf B}_g$ (\ref{gravebfield}) do not appear in (\ref{Gtt}, \ref{Gtj}, \ref{Gij}) as they did in (\ref{linmom}), nor in the energy equation (\ref{linE}). The gravito-electromagnetic fields seen in the force equation do not emerge ``naturally" in the field equations. We will go on to investigate the role of coordinate choices in the emergence of such effects.

The (\ref{Gtt}, \ref{Gtj}, \ref{Gij}) have profound implications for the dynamics of the fields. It follows from (\ref{Gtt}) that $\Psi$ obeys an elliptical equation, and from (\ref{Gtj}), so does $\bf w$. Taking the trace of (\ref{Gij}) reveals that $\phi$ also obeys an elliptical equation. In fact, it can be shown that only the traceless spatial components $s_{ij}$ of $h_{\mu\nu}$ have radiative character. The other components of $h_{\mu\nu}$ are determined entirely by the matter boundary conditions and by the $s_{ij}$, and their field equations merely provide initial value constraints. 

But we have left an important point unaddressed so far. It is that there are only 6 native gravitational degrees of freedom in the 10 components $h_{\mu\nu}$, and 6 independent equations. The other 4 equations are provided by the choice of coodinates. In general, coordinate information is mixed into the gravitational degrees of freedom. Therefore the choice of coordinates can affect the form of the field equations, and we will see how Maxwellian field equations arise from (\ref{Gtt}), (\ref{Gtj}), (\ref{Gij}) in two different coordinate systems, with new results in transverse coordinates compared to traditional results in harmonic coordinates.
\section{5. Linear field equations in transverse coordinates}

Various textbooks provide an analysis of the degrees of freedom in the linear gravitational field \cite{carroll, pw}, and we will not repeat those calculations. We instead summarize the main results for Maxwellian applications.

In a general coordinate system, independent degrees of freedom of the gravitational field can span multiple components of the metric, and the mixing among those components depends on the coordinate choice. Yet one can construct combinations of the metric potentials which are invariant under a linear coordinate transformation. Identification of the linear coordinate-invariant potentials is based on decomposing the $h_{tj}$ and $h_{ij}$ into irreducible transverse and longitudinal pieces and investigating their behavior under a coordinate transformation.\cite{pw,fh} When this is done, the 6 degrees of freedom of the linear gravitational field are exposed, without any coordinate information encoded. 

There is a unique coordinate system in which the degrees of freedom of the gravitational field map directly to components of the metric. In this coordinate system, the field equations take the same exact form as the coordinate-invariant field equations. Therefore, by working in this coordinate system, we can replicate the form of the coordinate-invariant field equations and expose the true degrees of freedom of the linear gravitational field. This special coordinate system is the transverse coordinate system, given by the 4 transverse conditions on the spatial components: 
\be 
\label{tg}
\p_i w^i =0\quad , \quad \p_i s^{ij} =0
\ee
Let us summarize the 6 native degrees of freedom of the linear gravitational field in this coordinate system.

Two of the 6 degrees of freedom are scalars. One is $\phi$, corresponding to the perturbation component $h_{tt}$, and representing the scalar of Newtonian gravity. The second scalar emerges from the trace of the spatial components, $\Psi$.

Two of the 6 degrees of freedom are components of the 3-vector $\bf w$, under the constraint that $\p_i w^i = 0$.

The other two degrees of freedom are components of the traceless part of the spatial components, $s_{ij}$, under the constraint (\ref{tg}), so these components are called ``transverse traceless" degrees of freedom. Including $\Psi$, there are 3 degrees of freedom in $h_{ij}$. These relationships are depicted in Figure 2.

\begin{figure*}
\includegraphics[width=1.0\textwidth]{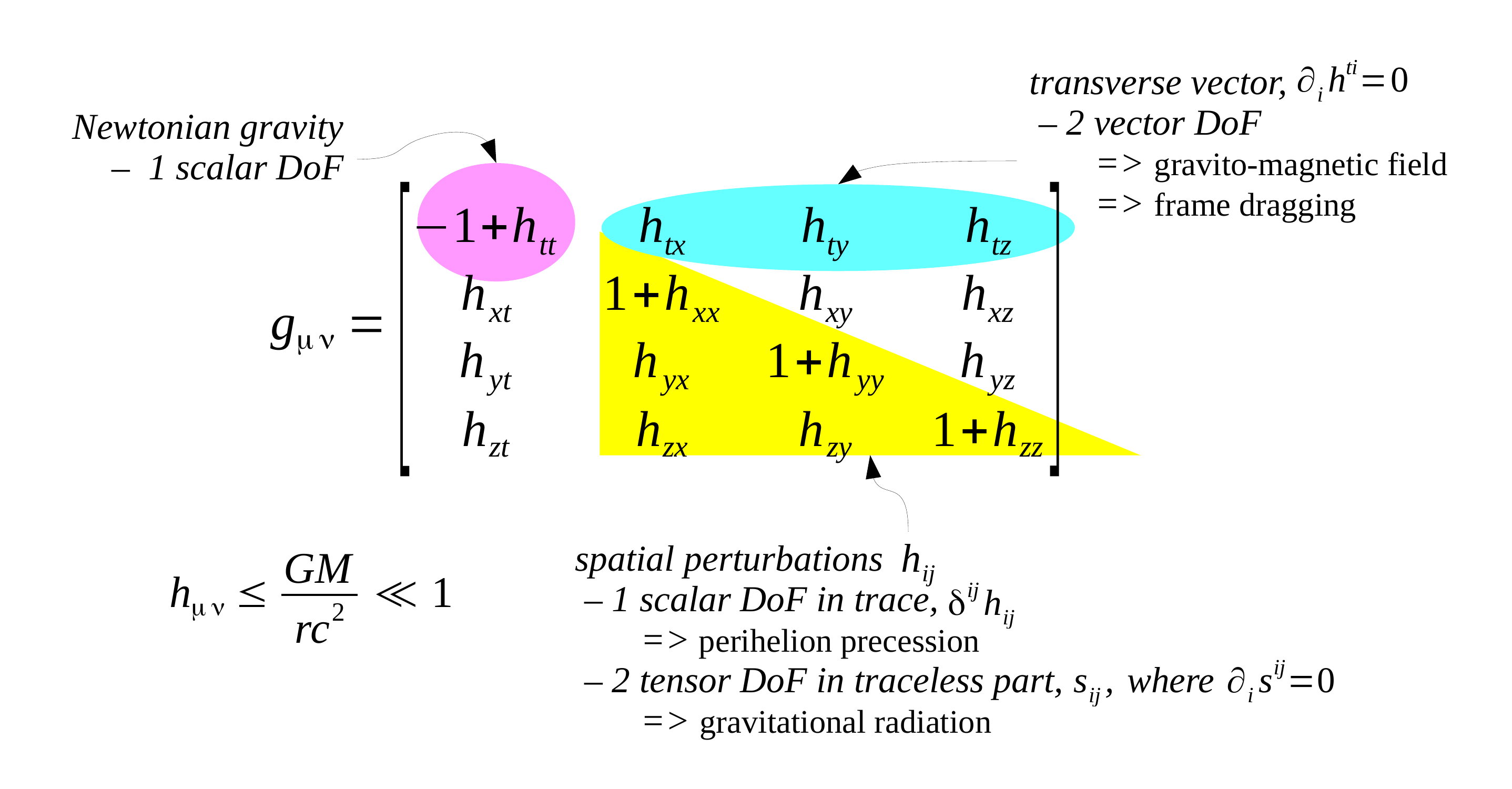}
\caption{Six native degrees of freedom (DoF) of the linear gravitational field, shown in transverse coordinates. Components of the metric have different physical effects, and can be classified according to how they behave under a spatial rotation. Newtonian gravity is located in the purple circle and behaves like a scalar under rotations; gravitomagnetism is in the blue oval and behaves like a vector under rotations; the spatial components in the yellow triangle behave like a tensor under rotations, and contain gravitational waves. The perturbations are all much less than 1; at the earth surface, the Newtonian potential $h_{tt} \sim 10^{-10}$. The two scalar degrees of freedom enter at zeroth order in the source velocity; the vector DoFs at first order in source velocity; and the tensor DoFs at second order in source velocity. The components $h_{\mu\nu}$ are decomposed in (\ref{lintime}) and (\ref{hij}).\label{Fig2}}
\end{figure*}

Now we can cast the Einstein equations in terms of the native potentials of transverse coordinates \cite{carroll, pw}
\be 
\label{EEtt}
2 \nabla^2 \Psi = {8\pi G\o c^4} T_{tt}
\ee
\be
\label{EEtj}
- {1\o 2} \nabla^2 w_j + {2\o c} \p_{tj} \Psi = {8\pi G\o c^4} T_{tj}
\ee
\be
\label{EEij}
(\delta_{ij} \nabla^2 - \p_{ij})(\phi - \Psi) - {1\o 2c}(\p_{ti} w_j + \p_{tj} w_i) +
{2\o c^2} \delta_{ij}\p_{tt} \Psi - \Box s_{ij} = {8\pi G\o c^4} T_{ij}
\ee

A simple scalar equation can be obtained by taking the trace of (\ref{EEij})
\be
\label{EEtrace}
2\nabla^2 (\phi -\Psi) + {6\o c^2} \p_{tt} \Psi = {8\pi G\o c^4} \delta^{ij} T_{ij}
\ee

Equations (\ref{EEtt}), (\ref{EEtj}), (\ref{EEij}), (\ref{EEtrace}) have the form of the coordinate-invariant field equations of the linear gravitational field.\cite{pw,fh} In a general coordinate system, the coordinate-invariant potentials will mix among components of $h_{\mu\nu}$. In this coordinate system, the coordinate-invariant potentials are individual components of $h_{\mu\nu}$.

Only the two degrees of freedom in $s_{ij}$ are radiative. In fact, the radiative degrees of freedom are in the transverse traceless spatial components in any coordinate system. Indeed, these are the components of the linear gravitational field that support gravitational waves. There are no propagating waves in the 4-vector part of the metric $h_{t\mu}$ that contains ${\bf E}_g$ and ${\bf B}_g$, as in (\ref{potwave}).

It is also interesting that the time-time component of energy-momentum, $T_{tt}$, drives $\Psi$ in (\ref{EEtt}). The Newtonian scalar $\phi$ is separately determined by (\ref{EEtrace}), and is equal to $\Psi$ in the Newtonian limit.

Refs.~\cite{pw}, \cite{fh} go further than we show here by decomposing the energy-momentum tensor into irreducible longitudinal and transverse components, and using constraints from the Bianchi identities. This results in a direct mapping of degrees of freedom of the gravitational field to degrees of freedom in the matter sources, and some further simplification of (\ref{EEtt}), (\ref{EEtj}), (\ref{EEij}).

Let us pursue the ${\bf E}_g$ and ${\bf B}_g$, (\ref{gravebfield}), into the field equations in transverse coordinates. We use the field equations (\ref{EEtt}), (\ref{EEtj}), (\ref{EEij}), (\ref{EEtrace}) to form the divergence and curl of ${\bf E}_g$ and ${\bf B}_g$ to find:
\be
\label{maxBtv}
 \bold{\nabla\times} {\bf E}_g = - {\p_t} {{\bf B}_g}/c\quad,\quad \bold{\nabla\times} {\bf B}_g = {16\pi G\o c^4} T_{tj} + {4\o c} \p_{t}{\bf N}_g
\ee
\be
\label{maxEtv}
\bold{\nabla\cdot} {\bf E}_g = -{4\pi G\o c^4}( T_{tt} + T_{ij}\delta^{ij}) + {3\o c^2} \p_{tt}\Psi \quad,\quad \bold{\nabla\cdot} {\bf B}_g = 0
\ee

The (\ref{maxBtv}) and (\ref{maxEtv}) are six equations in the components of two 3-vectors, corresponding to 4 equations in the 10 potentials. They combine information from (\ref{EEtt}), (\ref{EEtj}), (\ref{EEtrace}). The other 6 equations obtain from (\ref{EEij}). Note there is no gravitational displacement current, unless it is identified with ${\bf N}_g$. The absence of a $\p_t {\bf E}$ term in the gravitational Ampere law means there are no wave solutions for the gravito-electromagnetic 3-forces ${\bf E}_g$ and ${\bf B}_g$. We can add an additional set of 3-vector equations for ${\bf N}_g$ from (\ref{EEtt}):
\be
\label{maxNtv}
\bold{\nabla\cdot} {\bf N}_g = -{4\pi G\o c^4}T_{tt}
\quad ,\quad
\bold{\nabla\times} {\bf N}_g = 0
\ee

For the case of time-independent fields, the effects of $h_{ij}$ vanish and (\ref{maxBtv}) and (\ref{maxEtv}) reduce to the form of the time-independent Maxwell equations.

\section{6. Linear field equations in harmonic coordinates}

As discussed previously, the Einstein equations provide only 6 independent equations in the 10 unknowns of the metric components, and the other 4 are provided by the coordinate choice. 
There is a choice of coordinates which results in an extraordinary simplification of the linear Einstein equations (\ref{Gtt}), (\ref{Gtj}), (\ref{Gij}), the choice of harmonic coordinates:
\be
g^{\mu\nu}\Gamma^\a_{\mu\nu} = 0
\ee
which implies to first order that
\be
\label{hgauge}
2\p_\mu h^\mu_\nu = \p_\nu (\eta^{\a\b}h_{\a\b})
\ee

In this coordinate system, the field equations take the simple form \cite{weinberg}
\be
\label{harmonic}
\Box h_{\a\b} = -{16\pi G\o c^4}\l(T_{\a\b} - {1\o 2}\eta_{\a\b} T_{\mu\nu}\eta^{\mu\nu}\r)
\ee

Let us define the trace of the perturbations $h\equiv \eta^{\mu\nu}h_{\mu\nu}$. Then take the trace of (\ref{harmonic}) to find
\be
\Box h = - {16\pi G\o c^4} T_{\mu\nu}\eta^{\mu\nu}
\ee
This result can be used to rewrite the harmonic field equations (\ref{harmonic}):
\be
\label{hfe}
\Box\l( h_{\a\b} - {1\o 2} \eta_{\a\b}h\r) \equiv \Box H_{\a\b} = -{16\pi G\o c^4}T_{\a\b}
\ee
with the associated coordinate condition (\ref{hgauge}) rewritten:
\be
\label{hcc}
\p_\mu H^{\mu\nu} = 0
\ee
The coordinate conditions (\ref{hcc}) impose a constraint on the second derivatives,
\be
\label{hccdd}
\p_{tt} H^{tt}/c^2 = \p_{ij}H^{ij}
\ee
In harmonic coordinates, double time derivatives can be transformed into Laplacian operators. The relation (\ref{hccdd}) allows us to understand why hyperbolic field equations in these coordinates can be a mirage in the coordinates masking ``true" elliptic behavior in the degrees of freedom of the gravitational field.

Let us now break out the space and time components, and write the field equations for the 4 time components, along with their coordinate condition:
\be
\label{hve}
\Box H_{\a t} = -{16\pi G\o c^4}T_{\a t} \quad , \quad \p_\mu H^{\mu t} = 0
\ee
Here is almost an identity with (\ref{potwave}).

Likewise, we can write the field equations for the spatial components, along with their coordinate condition:
\be
\label{hsfe}
\Box H_{ij} = -{16\pi G\o c^4}T_{ij} \quad , \quad \p_\mu H^{\mu j} = 0
\ee

Now let us see how we can define Maxwell-like fields that incorporate the spatial components via $H_{\alpha t}$. Let us map (\ref{hve}) to (\ref{potwave}) under the electromagnetic gauge condition $\p_\mu A^\mu=0$, to deduce the form of Maxwell-like gravito-electric and gravito-magnetic fields in harmonic coordinates. We can write directly:
\bea
\label{harmE}
{\bf E}_h &\equiv&  \p_i H_{tt} - \p_t H_{ti}\nonumber \\
&=& \p_i h_{tt} -\eta_{tt}\p_i h/2 - \p_t h_{ti} \nonumber \\
&=& -\p_i\phi -\p_t {\bf w} - 3 \p_i\Psi \nonumber \\
&=& {\bf E}_g + 3 {\bf N}_g   
\eea
The gravito-electric field ${\bf E}_h$ in harmonic coordinates has an extra term in the gradient of the trace $h=-h_{tt}+\delta^{ij}h_{ij} = 2\phi - 6\Psi$ that is not in the gravito-electric field (\ref{gravebfield}) identified from the momentum equation, and this can be related to the neutral field, ${\bf N}_g$. In this way, the 3-vector ${\bf E}_h$ is constructed from 7 of the 10 components of $h_{\mu\nu}$: $h_{t\mu}$ and $h_{ii}$. The gravito-electromagnetic fields identified from the force equation involve only the 4 components $h_{t\mu}$ of the metric perturbation. The 3-vector gravito-electric field in harmonic coordinates in some sense represents the tensor nature of the gravitational field better than ${\bf E}_g$ from the force equation. 

Recall that the $h_{t\mu}$ of ${\bf E}_g$ and ${\bf B}_g$ approximates a 4-vector. While the $H_{t\mu}$ do involve 4 numbers, they are not a vector, nor do they approximate one. Consider the transformation of $H_{t\nu}$:
\be
\label{trev}
{H'}_{t\mu} = {\p {x}^\a \o \p t'}{\p {x}^\b \o \p {x'}^\mu} h_{\a\b} - {1\o 2}\eta_{t\mu} 
\delta^{\a\b} h_{\sigma\rho}{\p {x}^\sigma \o \p {x'}^\a}{\p {x}^\rho \o \p {x'}^\b}
\ee
The quantity (\ref{trev}) does not transform like a vector or a tensor. It is a mix of tensor potentials. 

The gravito-magnetic field is the same as ${\bf B}_g$:
\bea
\label{harmB}
{\bf B}_h &\equiv& \p_i H_{tj} - \p_j H_{ti}\nonumber \\
&=& \p_i  h_{tj} - \p_j h_{ti} \nonumber \\
&=& \nabla\times{\bf w} = {\bf B}_g
\eea
The ${\bf E}_h$ and ${\bf B}_h$ definitions imply a gravitational Faraday law between them, just as we saw for the definitions extracted from the force equation.

Now we can express the 4 field equations (\ref{hve}) in terms of the harmonic coordinate fields ${\bf E}_h$ and ${\bf B}_h$:
\be
\label{maxBbar}
\bold{\nabla\cdot} {\bf B}_h = 0 \quad,\quad \bold{\nabla\times} {\bf B}_h = {16\pi G\o c^4} T_{ti} +  \p_t {\bf E}_h/c
\ee
\be
\label{maxEbar}
\bold{\nabla\cdot} {\bf E}_h = -{16\pi G\o c^4} T_{tt} \quad,\quad \bold{\nabla\times} {\bf E}_h = - \p_t {\bf B}_h/c
\ee

The equations (\ref{maxBbar}) and (\ref{maxEbar}) are clearly Maxwellian in form. Yet the above considerations imply that the Maxwell-like equations (\ref{maxBbar}) and (\ref{maxEbar}) are underdetermined for the potentials they contain. It is necessary to employ the 6 spatial equations and their coordinate conditions, (\ref{hsfe}). 

If non-relativistic approximations to the source terms are taken, then Faraday induction is lost, as shown in the next section.

The equations (\ref{maxBbar}) and (\ref{maxEbar}) will lead to apparent wave equations in the ${\bf E}_h$ and ${\bf B}_h$ as in (\ref{maxwave}). According to our previous treatment in transverse coordinates, there is no radiative character in any of the 7 potentials comprising $H_{t\mu}$. However, the harmonic coordinate conditions (\ref{hcc}) mix the potentials among one another. The result is that ${\bf E}_h$ comprises any 7 of the 10 components of $h_{\mu\nu}$, not just the 7 components of $H_{t\mu}$ shown in (\ref{harmE}). Therefore we can rewrite ${\bf E}_h$ purely in terms of $h_{ij}$, using the coordinate condition of (\ref{hsfe}):
\be
{\bf E}_h \equiv  \p_i H_{tt} - \p_t H_{ti} = \p_i H_{tt} - \p_j H_{ij} = -4 \p_i \Psi - 2 \p_i s_{ij}
\ee
${\bf E}_h$ appears to have a radiative nature because it contains the $s_{ij}$, which in turn contain the radiative degrees of freedom of the linear gravitational field. However, as previously discussed, only the transverse traceless part of $s_{ij}$ contains the radiative degrees of freedom. Therefore $\p_i s_{ij}$ does not contain radiative degrees of freedom.

Likewise, the time-dependence of ${\bf B}_h$ can be expressed in terms of $s_{ij}$ by using the coordinate condition of (\ref{hsfe}):
\be
\p_t {\bf B}_h = \epsilon_{ijk}\p_j \p_t H_{kt} = \epsilon_{ijk}\p_{j}\p_m H^{mk} = \epsilon_{ijk}\p_{j}\p_m s^{mk}
\ee
Here, too, we see the $s_{ij}$ mixed into the harmonic gravito-magnetic field, and so its time-dependence appears to reflect a radiative nature. However, there are no radiative degrees of freedom in $\p_m s^{mk}$.

The force equation in harmonic coordinates is obtained from (\ref{linmom}) by using (\ref{harmE}) and (\ref{harmB}). Let us further restrict ourselves to the form correct to first order in particle speed, (\ref{linmom1}), and express it in terms of ${\bf E}_h$ and ${\bf B}_h$:
\bea
\label{linmom1h}
{d {\bf p}\o dt} &\simeq& m(
{\bf E}_h c^2 - 3{\bf N}_g c^2 + {{\bf v}} \times {\bf B}_h c
+ {2{\bf v}}\p_t\Psi 
- {2v^j}\p_t s_{ij} )  \\
&=& m( {\bf N}_g c^2 -2 \p_j s^{ij} c^2 + {{\bf v}} \times {\bf B}_h c+ {2{\bf v}}\p_t\Psi 
- {2v^j}\p_t s_{ij} )
\eea

We see that in spite of apparent similarities between some components of the gravitational field equations and the Maxwell equations, the force equation in these Maxwell-like coordinates involves many non-Lorentz terms in the spatial potentials $h_{ij}$. When everything is converted to the spatial potentials, the effects of ${\bf E}_h$ actually vanish! This means that the harmonic gravito-electromagnetic fields have a distinct Maxwellian form because they incorporate the spatial components of $h_{\mu\nu}$. However, the force equation in these terms is strikingly different from the Lorentz force law. Therefore, the price paid to work in harmonic coordinates and gain Maxwell-like equations, is to lose the classical form for the Lorentz force at lowest order in test particle speed. Even in the time-independent case, terms in $\Psi$ remain. So here is another way that the Maxwellian character implied in (\ref{maxBbar}) and (\ref{maxEbar}) is illusory: the ${\bf E}_h$ and ${\bf B}_h$ do not account for all the forces. 

\section{7. Harmonic coordinates to first order in source velocity}

It is common in harmonic coordinates to consider non-relativistic sources. For sources that have little internal energy, the energy momentum $T_{\mu\nu}\propto \rho U_\mu U_\nu$, where $\rho$ is the source mass density and $U_\mu$ is the covariant source velocity. Then if $U^i\ll c$,  we can ignore terms quadratic in source speed and approximate $T_{ij}\simeq 0$. 

This approximation also has implications for time derivatives of the fields. For solar system phenomena, we can assume the time-dependence of a field is due to motion of the source, so that $\p_t \phi /c \sim U^i \p_i \phi /c \ll \p_i \phi$. Time derivatives are therefore first-order corrections to the spatial gradients.

Yet we keep terms in $T_{tj}$ which are first order in the speed of the sources. This implies from (\ref{hsfe}):
\be
\label{hsfe0}
\Box H_{ij} = 0 \quad \rightarrow \quad h_{ij} = {1\o 2}\delta_{ij}h
\ee
That is, the spatial components $h_{ij}$ of the perturbation do not vanish, even if the spatial components $T_{ij}$ do. This is to be expected, from the Newtonian line element in the presence of static sources, (\ref{nle}). 

Now the coordinate condition of (\ref{hsfe}) implies $\p_t {\bf w} = 0$, so that $\p_t {\bf B}_h =0$. Under these approximations, the Maxwell-like equations (\ref{maxBbar}) and (\ref{maxEbar}) reduce to
\be
\label{Bnh}
\bold{\nabla\cdot} {\bf B}_h = 0 \quad,\quad \bold{\nabla\times} {\bf B}_h =- {16\pi G\o c^3} \rho {\bf U} + \p_t {\bf E}_h/c
\ee
\be
\label{Enh}
\bold{\nabla\cdot} {\bf E}_h = -{16\pi G\o c^2} \rho \quad,\quad \bold{\nabla\times} {\bf E}_h = 0
\ee
augmented with the coordinate conditions
\be
\label{hncc}
\p_t {\bf E}_h/c = -\nabla(\nabla\cdot{\bf w}) \quad, \quad \p_t {\bf B}_h = 0
\ee
Note that in this approximation, there is no Faraday induction. 

Also note that ${\bf B}_h \sim {\bf E}_h U/c \ll {\bf E}_h$. This means the gravito-magnetic force term is actually quadratic in small quantities, $vU/c^2$, so the proper force equation in this limit is
\be
\label{linmom1h2}
{d {\bf p}\o dt} =
mc^2({\bf E}_h  - 3{\bf N}_g)
 + O(\epsilon^2) = mc^2 {\bf N}_g + O(\epsilon^2)
\ee
where we are ignoring quantities quadratic in the small quanitity $\epsilon \sim U/c \sim v/c$. 

\section{8. Transverse coordinates to first order in source velocity}
Let us conclude by considering the Maxwellian field equations in transverse coordinates, (\ref{maxBtv}), (\ref{maxEtv}), (\ref{maxNtv}), ignoring terms of second order in source velocity. Equation (\ref{EEtrace}) implies
\be
\label{EEtracea}
\nabla^2 (\phi - \Psi)  \simeq 0 \quad \rightarrow \quad \phi \simeq \Psi
\ee
Equation (\ref{EEij}) implies
\be
\label{EEij1}
 - {1\o 2c}(\p_{ti} w_j + \p_{tj} w_i) 
\simeq \nabla^2 s_{ij}
\ee
Equations (\ref{maxBtv}), (\ref{maxEtv}) reduce to
\be
\label{maxBtv1}
 \bold{\nabla\times} {\bf E}_g \simeq 0\quad,\quad \bold{\nabla\times} {\bf B}_g = -{16\pi G\o c^3} \rho {\bf U} + {4\o c} \p_{t}{\bf N}_g
\ee
\be
\label{maxEtv1}
\bold{\nabla\cdot} {\bf E}_g \simeq -{4\pi G\o c^2}\rho  \quad,\quad \bold{\nabla\cdot} {\bf B}_g = 0
\ee
There is no gravitational Faraday law at this level of approximation. The equations (\ref{maxNtv}) stay the same.

The force equation in this limit reduces to Newton's law:
\be
{d {\bf p}\o dt} \simeq mc^2 {\bf E}_g
\ee

\section{9. Key Maxwellian gravity works in perspective}
\subsection{9.1 Sciama on inertia}
Sciama's objective \cite{sciama} was to show inertia could arise from a gravitational inductive effect with the mass of the universe. He did not assume the existence of a comprehensive set of vector gravity equations like (\ref{maxBbar}) and (\ref{maxEbar}). Instead, he assumed key elements of them, on the grounds that such effects were likely to exist in more-complicated theories of gravity. He assumed the expressions (\ref{gravebfield}), assumed a solution to the inhomogeneous equations of (\ref{maxBbar}) and (\ref{maxEbar}), and then made the essential connection between the speed of the source in $T_{tj}\propto U_j$ with the speed of an accelerated object. Mathematically, he assumed that in the instantaneous rest frame of an accelerated body, the gravitational field of the universe is zero, and then formulated this statement in terms of the induced gravito-electric field.

His positive identification of inertial induction in vector gravity was grounds to look for the same effect in tensor gravity, but it did not prove the effect exists in tensor gravity. Indeed, Ref.~\cite{cw} devoted an entire book to the attempt to replicate the inertial induction result in tensor gravity that Sciama obtained simply in vector gravity. They did not succeed particularly, and could only argue that inertia could be understood in tensor gravity as a gravito-magnetic effect, seen through the prism of the Kerr metric in Boyer-Lundquist coordinates.

Sciama used the time-independent solutions to the elliptic equation for $\bold w$, as in the time-independent version of (\ref{Bnh}). However, he combined this with the Faraday's Law that results from $\nabla\times {\bf E}_g = - \p_t {\bf B}_g$. Therefore, Sciama's inertial induction effect does appear to be a component of the linearized tensor force in (\ref{linmom}), and it might be possible to explain inertia within general relativity by focusing on $\bold w$. Yet there may be an inconsistency in dropping the displacement current in (\ref{Bnh}) while keeping the time derivative of ${\bf w}$ in the definition of ${\bf E}_h$.

Another problem with the Sciama approach is that the potentials from the mass of the universe are not small, they are large. So the assumption of linear gravity is broken when the mass of the universe is considered. While Sciama showed that vector gravity can explain inertia as a gravitational interaction with the mass of the universe, it does not mean that inertia would arise from tensor gravity through the same mechanism. The question of Sciama-style inertial induction remains open in general relativity.

\subsection{9.2 Forward on experimental gravity}
Forward \cite{forward} was motivated to provide theoretical support to experimental efforts to measure effects of general relativity. Forward did not write down a set of vector gravity equations like (\ref{maxBbar}) and (\ref{maxEbar}), but he did write their solutions. His use of vector Maxwellian gravity was discriminating, employing differing considerations to different components of the metric.

The crux of Forward's approach was to use harmonic coordinates to derive the apparent hyperbolic equation in the perturbations, (\ref{hfe}). The similarity to (\ref{potwave}) compelled the identification of analogies with electromagnetism, regarding gravito-electric and gravito-magnetic forces, and mass currents. Forward generally adopted the assumptions of time-independence and non-relativistic motion of bodies. Because of this, he effectively solves for the potentials under the equations (\ref{Bnh}), (\ref{Enh}), (\ref{hncc}). Both $\phi$ and $\bold w$ are governed by elliptic equations, as they should be.

The implied momentum equation at this order of approximation is (\ref{linmom1h2}), yet Forward used (\ref{linmom1}) with the time-dependent terms truncated, to look like the Lorentz force law. His expression for the electric field included the time derivative term, ${\bf E} = -\nabla \phi - \p_t \bold w$, which implied Faraday induction, in contradiction to (\ref{Enh}). So here is a common mistake in the gravito-electromagnetic literature: to formulate the field equations and force equations inconsistently in terms of their ordering in small quantities.

Forward was aware of the importance of $h_{ij}$ in solar system phenomena, particularly perihelion precession, for which he provided a calculation. And he was aware there was no analog for these components in vector gravity. Therefore, he was not bound by vector gravity in the entirety of his paper, it was just a tool of opportunity. He also linearized the Schwarzschild metric as the basis for force calculations.

Forward erroneously assumed that (\ref{hfe}) implied waves in all components $h_{\mu\nu}$. This misconconception was understandable in 1960, but it has remained a consistent misconception throughout the literature since then. Forward recognized the proper harmonic coordinate conditions, but did not seem to understand how they would constrain propagating degrees of freedom to only the $s_{ij}$ components of $h_{\mu\nu}$.

Nonetheless, Forward's calculation of the force of frame-dragging from the gravito-magnetic effect has merit. The gravito-magnetic force is perhaps the only electromagnetic analogy that has independent existence, and it should be an important non-Newtonian force in steady-state astrophysical situations. Yet a study of gravito-magnetism without induction might be unsatisfactory.
\subsection{9.3 Harris and BCT on Maxwellian gravity}
Harris developed Maxwellian gravity through harmonic coordinates, under the condition that terms second order in source velocity are ignored. Therefore, the Harris field equations (16) of that paper match our (\ref{Bnh}) and (\ref{Enh}).

Harris tied his results to earlier Maxwellian results by Braginsky, Caves, \& Thorne (BCT) \cite{bct}. They proceeded from a PPN framework, but in fact their results match our transverse coordinate results. Specifically, equations (3.8) of \cite{bct} are of the same form as our (\ref{maxBtv}), (\ref{maxEtv}).

Yet both Harris and BCT set $s_{ij}\rightarrow 0$ at the outset, without discussion. Therefore, both are restricted to non-radiative modes of the gravitational field. They both also equate the spatial diagnonal metric perturbations to the time-time component, as found for the Newtonian limit (\ref{nle}). Therefore, both Harris and BCT model the gravitational field with 4 potentials. This creates a degeneracy in the distinction between the ${\bf E}_h$ used by Harris and the ${\bf E}_g$ used by BCT. The former involves the spatial diagonal components of the metric, but they are assumed equal to the time-time component, thereby reducing ${\bf E}_h$ in Harris to consisting of 4 potentials.

The PPN approach created some confusion for Harris, because the field equations are of course different for the ${\bf E}_g$ and ${\bf E}_h$. Harris rightly calculated that $\nabla\times {\bf E}_h =0$ per (\ref{Enh}) for approximations to first order in source velocity. Yet he was puzzled that BCT retained the induction term. In fact, Faraday induction still vanishes in a consistent ordering to first order in source velocity, as shown in our (\ref{maxBtv1}). 

BCT expressed their force equation in terms of relativistic momentum, as we do here. Harris worked in terms of coordinate acceleration. Nonetheless, Harris, like Forward, used inconsistent ordering in the force equation and field equations. At the order of approximation used by Harris in the field equations, the proper force equation is (\ref{linmom1}), yet Harris used a truncated version of it. BCT captured effects from ${\bf N}_g$ in their force equation, yet they, too appeared to neglect time derivative terms appearing in (\ref{linmom1}). So it appears Harris used the field equations in terms of harmonic fields ${\bf E}_h$, yet used the force equation in terms of ${\bf E}_g$. This appears to be a common mistake in the gravito-electromagnetic literature.

\section{10. Conclusions}
In the limit of linear tensor gravity, gravito-electric and gravito-magnetic forces exist in the gravitational force equation (\ref{linmom}). These gravito-electric and gravito-magnetic fields obtain from the potentials of a 4-vector-like quantity that is the $h_{t\mu}$ components of the metric perturbation, and it transforms approximately as a 4-vector. There is an inductive, Faraday relationship between the ${\bf E}_g$ and ${\bf B}_g$ of (\ref{gravebfield}).

Yet the gravitational force cannot be fully characterized with just the ${\bf E}_g$ and ${\bf B}_g$ force components. Spatial metric perturbations are generated by sources at rest. They can be physically important, such as for perihelion precession. But such perturbations are missing from a pure 4-vector-like description involving just 4 of the 10 potentials.

The gravito-electric and gravito-magnetic fields ${\bf E}_g$ and ${\bf B}_g$ identified in the linear force equations do not appear as naturally in the linear field equations. The field equations provide only 6 independent equations for the 10 components of the metric, and a coordinate choice is required to provide the other 4 equations to close the system. The freedom of the coordinate choice allows Maxwellian forms to manifest in the field equations, but the Maxwellian forms depend on the coordinate choice.

Modern analysis shows that the 6 degrees of freedom in the linear gravitational field can be identified by constructing coordinate-invariant potentials from among the 10 component potentials of $h_{\mu\nu}$. In transverse coordinates, the field equations for the $h_{\mu\nu}$ are identical in form to the coordinate-invariant field equations. Components of $h_{\mu\nu}$ then map directly to gravitational degrees of freedom. The 6 gravitational degrees of freedom are identified as two scalars, the Newtonian potential $\phi$, and a potential in the trace of the spatial components, $\Psi$; two components in the 3 vector ${\bf w}$, under the constraint that $\nabla\cdot{\bf w} =0$; and two components in the transverse traceless spatial components, $s_{ij}$. 

We obtained the gravitational Maxwell-type equations (\ref{maxBtv}), (\ref{maxEtv}), (\ref{maxNtv}) in transverse coordinates. We showed the importance of the neutral field ${\bf N}_g$ in these coordinates, a force 3-vector with no analogy in electromagnetism. 

It is clear in transverse coordinates that there are no gravitational waves -- no propagating degrees of freedom -- in the 3-vector gravito-electromagnetic fields ${\bf E}_g$ and ${\bf B}_g$. Nor are there wave solutions in the components of $h_{t\mu}$. The only freely-propagating degrees of freedom are in the transverse traceless spatial components $s_{ij}$. 

A great simplification of the linear tensor field equations can be obtained by working in harmonic coordinates. In these coordinates, components of the field equations assume Maxwellian forms in terms of harmonic gravito-electric and gravito-magnetic fields ${\bf E}_h$ and ${\bf B}_h$. Expressions in these coordinates form the basis for most work on gravito-electromagnetic fields.

The ${\bf E}_h$ and ${\bf B}_h$ are tensor expressions, involving 7 components of $h_{\mu\nu}$, and this coordinate system mixes independent degrees of freedom among individual components of the metric. The 6 harmonic Maxwell-like 3-vector field equations do not contain all the information to close the system of field equations; they are underdetermined for the 7 components of the perturbation they contain. The remaining 6 equations are provided by the coordinate conditions and by two spatial field equations.

In harmonic coordinates, the force equation does not have a Lorentz form. It involves gradients of the spatial metric components at lowest order, and there is no analogy for such effects in electromagnetism. Therefore harmonic coordinates yields Maxwellian forms in the field equations, but they do not correspond to Lorentz-like forms in the force equation.

In harmonic coordinates, the ${\bf E}_h$ and ${\bf B}_h$ appear to obey radiative, Maxwell-like equations. But when the coordinate conditions are applied, the radiative nature in the gravito-electromagnetic fields exists only through terms in $\p_i s^{ij}$, which carry no radiative degrees of freedom. 

These results are summarized in Table I, which compares gravito-electromagnetism defined in two coordinate systems. One chooses gravito-electromagnetic fields based on the simplification of the field equations arising in harmonic coordinates, and one chooses fields based on gravito-electromagnetic effects in the force equation.

We find that consistent ordering of terms in the field equations and in the force equation is often overlooked in conventional formulations of gravito-electromagnetism. An assumption of non-relativistic sources implies an absence of Faraday induction in the field equations, and implies gravito-magnetic forces are second order in small quantities.

There are two separate ordering parameters in linear gravitation: the speed of test particles, and the speed of gravitational sources. In the force equation, gravito-electric fields exist at zeroth order in test particle velocity, gravito-magnetic effects at first order, and gravito-neutral (tensor) effects at second order. Yet in the field equations, the electric field and the neutral field are both generated at zeroth order in source speed, while the magnetic field is first order in source speed. Pure tensor effects are generated at second order in source speed. Because time-dependence arises in the fields mainly from the motion of the sources, time-derivatives of the fields are corrections to the gradients at first order in source speed. Many treatments of gravito-electromagnetism will make inconsistent ordering choices between the field equations and force equations, or else truncate terms of relevant order from the force equation. Often such mistakes reflect an attempt to force exact Maxwellian analogs simultaneously in both the field equations and the force equation.

In summary, several mirages appear in the formulation of Maxwellian vector gravity by approximation from general relativity: 
\begin{itemize}
\item{3-vector gravito-electromagnetic waves are naively implied in harmonic coordinates, when such fields actually carry no radiative degrees of freedom}
\item{with a suitable choice of coordinates, Maxwellian forms can be obtained in the field equations, or Lorentz-like forms can be obtained at lowest order in the force equation, but no single choice of coordinates yields both Maxwellian field equations and a Lorentz-like force equation}
\item{the gravito-magnetic field in the force equation is actually second order in small quantities, but is often retained in the force law while other quantities at second order are neglected.}
\item{the gravitational Gauss and Faraday law are constrained together by ordering in source velocity, so that the Gauss source term must be carried to at least second order in source velocity to preserve the Faraday law. An approximation of non-relativistic sources necessarily sets induction to zero.}
\item{the gravito-electromagnetic fields defined from the force equation or from the harmonic coordinate choice are coordinate-dependent. In a subsequent paper, we will show how to construct coordinate-invariant gravito-electromagnetic fields.}
\end{itemize}

\begin{table*}
\caption{\label{tab:table1}Maxwellian gravito-electromagnetic forces and fields, in transverse and harmonic coordinates, for non-relativistic test bodies, and for relativistic and non-relativistic sources. The two coordinate choices correspond to two different identifications of the gravito-electromagnetic fields. All potentials are unitless, and all 3-forces have units of inverse length. Electric 3-forces can be converted to units of acceleration by multiplying by $c^2$. A mixed notation of bold type and small roman indices is used to indicate 3-vector components.}
\begin{ruledtabular}
\begin{tabular}{c|cc}
 &Transverse coordinates & Harmonic coordinates\\  \hline \\ 
Significance & Potentials $\rightarrow$ Degrees of freedom &Simple field equations  \\ \\
coordinate conditions &
\(\displaystyle\p_i w^i =0\quad , \quad \p_i s^{ij} =0 \)&
\(\displaystyle\p_\mu H^\mu_\nu = 0 \)
\\ \\
electric field & ${\bf E}_g \equiv -\nabla\phi -\p_t {\bf w}/c$ & 
${\bf E}_h \equiv \p_i H_{tt} - \p_t H_{ti} = {\bf E}_g + 3 {\bf N}_g$  \\ \\
magnetic field & ${\bf B}_g \equiv \epsilon_{ijk}\p_j w^k$ & 
${\bf B}_h \equiv \epsilon_{ijk}\p_j H_{tk} = {\bf B}_g $\\ \\
neutral field & ${\bf N}_g \equiv -\nabla\Psi$ & 
none defined\\ \\
E \& B definition& force equation &field equations \\ \\
DoF in E + B & 3 & 4 \\ \\
comps. of $h_{\mu\nu}$ in E + B & 4 & 7 \\ \\
non-relativistic & 
\(\displaystyle {1\o\varepsilon}{d {\bf p}\o dt} \simeq
{\bf E}_g + {{\bf v}\o c} \times {\bf B}_g
+ {2{\bf v}\o c^2}\p_t\Psi 
- {2v^j\o c^2}\p_t s_{ij} \)& 
\\ force equation & & \\
(first order in $v/c$)& &
\(\displaystyle{1\o\varepsilon}{d {\bf p}\o dt} \simeq
{\bf E}_h - 3{\bf N}_g + {{\bf v}\o c} \times {\bf B}_h
+ {2{\bf v}\o c^2}\p_t\Psi 
- {2v^j\o c^2}\p_t s_{ij} \)\\ \\
 & 
\(\displaystyle\bold{\nabla\times} {\bf E}_g = - {\p_t} {{\bf B}_g}/c \)&
\(\displaystyle\bold{\nabla\times} {\bf E}_h = - \p_t {\bf B}_h/c \)
 \\ electric field  &&\\
&
\(\displaystyle\bold{\nabla\cdot} {\bf E}_g = -{4\pi G\o c^4}( T_{tt} + T_{ij}\delta^{ij})+{3\o c^2} \p_{tt}\Psi \)&
\(\displaystyle\bold{\nabla\cdot} {\bf E}_h = -{16\pi G\o c^4} T_{tt} \)
\\ \\
&
\(\displaystyle \bold{\nabla\times} {\bf B}_g = {16\pi G\o c^4} T_{tj} + {4\o c} \p_{t}{\bf N}_g \) &
\(\displaystyle \bold{\nabla\times} {\bf B}_h = {16\pi G\o c^4} T_{tj} +  \p_t {\bf E}_h/c \)
\\ magnetic field && \\
&
\(\displaystyle \bold{\nabla\cdot} {\bf B}_g = 0 \)&
\(\displaystyle \bold{\nabla\cdot} {\bf B}_h = 0 \)
\\  \\ 
&
\(\displaystyle \bold{\nabla\times} {\bf N}_g = 0 \) &
not applicable
\\neutral field&& \\
&
\(\displaystyle \bold{\nabla\cdot} {\bf N}_g = -{4\pi G\o c^4}T_{tt} \) &
not applicable
\\  \\ 
spatial potentials &
cf. Eqn (\ref{EEij}) & 
\(\displaystyle\Box H_{ij} = {16\pi G\o c^4}T_{ij} \)
\\ \\
\hline \\
& 
\(\displaystyle
{d {\bf p}\o dt} \simeq mc^2 {\bf E}_g
\) &
\(\displaystyle{d {\bf p}\o dt} \simeq
mc^2({\bf E}_h  - 3{\bf N}_g) =  mc^2{\bf N}_g  \)
\\modifications for&& \\ 
non-relativistic sources&
\(\displaystyle\bold{\nabla\times} {\bf E}_g \simeq 0 \)&
\(\displaystyle\bold{\nabla\times} {\bf E}_h \simeq 0 \)
\\ to first order&&\\
in $v/c \sim U/c$&
\(\displaystyle\bold{\nabla\cdot} {\bf E}_g \simeq -{4\pi G\o c^2}\rho \)&
\(\displaystyle\bold{\nabla\cdot} {\bf E}_h = -{16\pi G\o c^2} \rho \)
\\ \\
 $T_{ij}\simeq 0$&
\(\displaystyle \bold{\nabla\times} {\bf B}_g = -{16\pi G\o c^3} \rho {\bf U} + {4\o c} \p_{t}{\bf N}_g \) &
\(\displaystyle \bold{\nabla\times} {\bf B}_h = -{16\pi G\o c^3} \rho {\bf U} +  \p_t {\bf E}_h/c \)
\\ $T_{t\mu}\equiv\rho c U_\mu$&&\\
&
\(\displaystyle - {1\o 2c}(\p_{ti} w_j + \p_{tj} w_i) 
= \nabla^2 s_{ij}, \quad\phi \simeq\Psi \) &
\(\displaystyle h_{ij} = {1\o 2}h \delta_{ij} \)
\\ \\

\end{tabular}
\end{ruledtabular}
\end{table*}

\section{Acknowledgments}

This work was supported by DARPA DSO under award number D19AC00020.

\end{document}